\documentclass[a4paper,12pt]{article}

\newcommand{\be}{\begin{equation}}
\newcommand{\ee}{\end{equation}}
\usepackage[dvips]{graphicx}

\begin{document}

{\centerline{\Large\bf Can flaring stars explain the annihilation line}}
 {\centerline{\Large\bf from the Galaxy bulge? }}

\vspace{1cm}

{\centerline{G.S.~Bisnovatyi-Kogan$^{1,2}$, A.S. Pozanenko$^{1,2}$}
\vspace{0.5cm}

$^1$Space Research Institute, Russian Academy of Sciences, Moscow,
Russia,

$^2$National Research Nuclear University MEPhI,
 Moscow, Russia}

\vspace{1cm}

 \noindent

\section{Introduction}

Electron-positron annihilation line from a Galactic center direction was discovered by the balloon-borne germanium gamma-ray telescope \cite{Leventhal_1978}, and  confirmed by OSSE experiment of CGRO mission \cite{Purcell_1993,Purcell_1997}. Extensive observations  by INTEGRAL observatory \cite{jean_2003,chur11} permit to determine properties of the annihilation line from Galactic bulge. The observations confirmed the steady state of the line, and estimate production rate of positrons ($e^{+}$) in the  bulge  as  $2 \times 10^{43}$  \cite{siegert_2015}. Possible sources of $e^{+}$  already discussed are supernovae explosions \cite{Prantzos_2006,Alexis_2014}, microquasars \cite{siegert_2016},   gamma-ray bursts \cite{parizot_2005}, tidal disruption events \cite{cheng_2006},  activity near black hole of Sgr A*, e.g. \cite{totani_2006}, $e^{+}$ generation by subrelativistic cosmic rays \cite{Pshirkov_2016}, or even dark matter, e.g. \cite{Ascasibar_2006}. One remarkable feature of the  line emission is an absence of resolved point like sources \cite{siegert_2015}. Any model should take into account this feature. We consider flares of low-mass stars as a possible cumulative $e^{+}$ source of the observed annihilation line from the bulge.  Our estimations show that $e^{+}$  production by numerous flaring stars in Galaxy bulge can be significant, or even the main source of $e^{+}$ responsible for the annihilation line observed from the central part of our Galaxy.

\section{Positrons in solar flares}

The number of positrons generated during a strong solar flare may be estimated from the intensity of the annihilation line  of $E_\gamma=511$ keV observed in powerful flares. It was determined that fluence of  $E_\gamma=511$ keV line over the entire  flare of 2002 July 23 (duration of observation is 960 s) was $\sim 83 \pm 14$ cm$^{-2}$ \cite{Share_2003}. The flare  was registered by  high purity Ge-detectors of RHESSI mission.  One can estimate the number of $e^{+}$ ejected during the flare provided isotropic line emission. The Sun-Earth distance is equal to $D=1.5\cdot 10^{13}$ cm, the total number of photons emitted during the flare in the 511 keV line   is equal to

\begin{equation}
 \label{ansun1}
N_{511,sun} =  F_{511,sun}4\pi D^2    = 2.3 \cdot 10^{29}.
 \end{equation}
%
%
%

Despite that annihilation line is a clear signature of a positron production, exact positron number in the flare is not known, at least, due to  following reasons. A ratio of  the two channels of annihilation  is  poorly understood. One of them assumes intermittent positronium production in a triplet spin state,  and further annihilation with an emission of 3 photons. The annihilation via this channel produces continuum emission but not the annihilation line of 511 keV. The ratio of annihilation rates of different channels  is known as the $3\gamma/2\gamma$ ratio.  Also the line emission could be non-isotropic because positron production related to a small region of flaring activity of the Sun surface. And finally, only a small number of positrons produced in flares can be annihilated while traveling to the Earth.

Assuming that only part $\delta$ of the emitted positrons  are annihilated, we can estimate a total  positron production in the flare as $N_{511,sun}=2.3\cdot 10^{29}/\delta\,\,$. The positrons produced in solar flares due to beta decay of radioactive isotopes are almost relativistic, and one of the signature of relativistic positron production is an observed  polarization of a microwave radiation in solar flares \cite{Fleishman_2013}. The cross-section of the annihilation \cite{blp71} is

\begin{equation}
 \label{annsect}
 \sigma_{ann} \approx \pi\,r_e^2=2.5\cdot 10^{-25} {\rm cm}^2,
\end{equation}
where $r_e=2.6\cdot 10^{-13}$ cm is the classical electron radius.

The electron number density in the solar corona may be approximated by a power law function \cite{scord},
with a density at the base $n_{cor,0}\approx 10^8$ cm$^{-3}$.
The number of electrons on the line of sight  from the Sun to the Earth is estimated as
$N{e,cor}\approx n_{cor,0}\, R_\odot\approx 7\cdot 10^{18}\, $cm$^{-2}$. It gives a probability of the
annihilation for the positrons emitted from the Sun as
$W_{ann} \approx \sigma_{ann} N_{e,corr}\, \approx \, 2\cdot 10^{-6} \sim \delta$.
For this value of $\delta$,  the number of the positron production in the flare as $r_{e,\odot,flare}=N_{511,sun}/\delta
=2.3\cdot 10^{29}/\delta \approx 1.2\cdot 10^{35}\,\, $. Actually the positron production  is two times less, because
during $e^+\, e^-$  annihilation two photons with energy 0.511 MeV are produced, so the number of positrons
is equal to

\begin{equation}
 \label{solar_positron_rate}
N_{e,\odot,flare}=0.5 \times N_{511,sun}/\delta
 =0.6\cdot 10^{35}\,\,.
\end{equation}
We can estimate now  $\eta$ the ratio of the positron production to a bolometric energy of the flare. For the X4.8 class flare described above the bolometric  energy is about $5\cdot 10^{32}~ erg$ \cite{Schrijver} and

\begin{equation}
 \label{conversion_rate}
 \eta = 0.6\cdot 10^{35}/5\cdot 10^{32} = 120 \,\,  e^{+}/erg
\end{equation}

\section{Positron rate from flaring stars in the bulge}

To estimate the positron rate production by flaring star in the bulge we need a flare frequency distribution, a number of flaring stars in the bulge, and the ratio of positron production per bolometric energy of the flare $\eta$ obtained above.  The flare frequency of flaring stars was investigated using data obtained by Kepler space-borne telescope \cite{Kitze_2014,Candelaresi_2014,Wu_2015}. The frequency distribution $\nu(E)$ of 4494 superflares detected from 77 stars of G-class is $\nu(E)= A\cdot E^{-\gamma}$, where ${\gamma}=2.04\pm0.17$. ($E$ is a bolometric energy of the flare.) Index  ${\gamma}$ is varying for individual stars from 0.65 up to 2.45.  For further estimations we suggest that  power law index ${\gamma}$ is equal to 2.  A typical  normalization factor for frequent flare activity stars observed by Kepler mission (see Fig. 8 of \cite{Wu_2015})  is $A\approx 10^{36} erg^{-1}year^{-1}$.

The total flare energy $E_{tot}$ per year per one flaring star can be written as

\begin{equation}
 \label{flare_energy_rate0}
 E_{tot} = \int  \nu(E)\cdot E dE, \,\, erg/year
\end{equation}

The frequency distribution is restricted by the energy of most intense flares in the Kepler sample, as $2 \cdot 10^{37}$ erg (see however  \cite{DG_CVn}). The low limit of flare energy cannot be determined due to insufficient sensitivity. However we can use extensive solar flare observations and put lower limit   $ 10^{24}$ erg, i.e. the same as observed for the solar flares \cite{Schrijver}. Using these  limits we obtain

\begin{equation}
 \label{flare_energy_rate}
 E_{tot}  =  A  \cdot \ln {(E_{max}/E_{min})}  \simeq 3\cdot 10^{37}  \,\, erg/year
\end{equation}

Accepting the mass of the bulge $M_b = 2\cdot 10^{10}\, M_\odot$,
\cite{Valenti_2015}
one can estimate the total number of stars of G,K,M - types as $N_{stars} = M_b / (0.3 \cdot M_\odot) \simeq 6\cdot 10^{10}$.
Then we can write a total production rate of positrons in the bulge $N_e^{+}$ as

\begin{equation}
 \label{positron_bulge_rate}
 N_{e^{+}} = N_{stars} \cdot E_{tot}  \cdot \eta  = 1.8\cdot10^{48} \eta \,\, e^{+}/year
\end{equation}
The estimated rate of $e^+ e^-$ annihilation in the bulge  is $R_{e^{+}} = 2 \cdot 10^{43}\, s^{-1} \approx 6 \cdot 10^{50}\, year^{-1} $ \cite{siegert_2015}.
Such rate of positron generation should be present at steady state. From (\ref{positron_bulge_rate}) it follows that bursting red dwarfs could supply the necessary
amount of positrons at $\eta=340$, what is about 3 times larger than the value of $\eta\simeq120$ estimated for the Sun (\ref{conversion_rate}).

 Some parameters used for estimating $N_{e^{+}}$ such as minimal and maximal  energy of flares, conversion coefficient $\eta$ and the index $\gamma$ are poorly investigated and by varying these parameters one can increase the total $e^{+}$ production rate approaching to the observed value. For the solar flare mentioned above \cite{Share_2003} only $10^{-4}$ of bolometric energy of the flare was used for the positron production. For more powerful flares  this value should be at least 3 times larger to satisfy the estimated $e^{+}$ production  obtained from observations of the annihilation line. Crude estimating show  that positron production by flaring stars can be significant, or even the main source of positrons responsible for the annihilation line observed from the central part of our Galaxy.

 Our hypothesis could be verified by searching for annihilation line in nearest and massive globular clusters enriched by red-dwarfs.  The brightest globular cluster NGC 5139 at the distance of 4.8 kpc and the mass of $\sim 4\cdot 10^{6}\, M_\odot$ \cite{Ven_2006} can be considered. A flux ratio between observed  $e^{+}e^{-}$ annihilation line from Galaxy bulge and assumed flux of NGC 5139 should be  $\sim 1.8\cdot 10^{3}$. Observation with necessary sensitivity could be available in future.


\begin{thebibliography}{99}


\bibitem{Leventhal_1978} 	Leventhal, M., MacCallum, C. J.,  Stang, P. D. Detection of 511 keV positron annihilation radiation from the galactic center direction. ApJ, \textbf{225}, L11 (1978).


\bibitem{Purcell_1993} Purcell, W. R.; Grabelsky, D. A.; Ulmer, M. P.; Johnson, W. N.; Kinzer, R. L.; Kurfess, J. D.; Strickman, M. S.; Jung, G. V. OSSE observations of Galactic 511 keV positron annihilation radiation - Initial phase 1 results. ApJ, \textbf{413}, L85 (1993).

\bibitem{Purcell_1997} Purcell, W. R.; Cheng, L.-X.; Dixon, D. D.; Kinzer, R. L.; Kurfess, J. D.; Leventhal, M.; Saunders, M. A.; Skibo, J. G.; Smith, D. M.; Tueller, J. OSSE Mapping of Galactic 511 keV Positron Annihilation Line Emission. ApJ, \textbf{491}, 725 (1997).

\bibitem{jean_2003} Jean, P.; Knödlseder, J.; Lonjou, V.; Allain, M.; Roques, J.-P.; Skinner, G. K.; Teegarden, B. J.; Vedrenne, G.; von Ballmoos, P.; Cordier, B.; Caraveo, P.; Diehl, R.; Durouchoux, Ph.; Mandrou, P.; Matteson, J.; Gehrels, N.; Sch\"{o}nfelder, V.; Strong, A. W.; Ubertini, P.; Weidenspointner, G.; Winkler, C.
  A\&A \textbf{407}, L55 (2003).

\bibitem{chur11} Churazov E., Sazonov S., Tsygankov S., Sunyaev R., Varshalovich D.
   Positron annihilation spectrum from the Galactic Centre region observed by SPI/INTEGRAL revisited: annihilation in a cooling ISM?    MNRAS \textbf{411}, 1727 (2011).

\bibitem{siegert_2015} Siegert, Thomas; Diehl, Roland; Khachatryan, Gerasim; Krause, Martin G. H.; Guglielmetti, Fabrizia; Greiner, Jochen; Strong, Andrew W.; Zhang, X. Gamma-ray spectroscopy of positron annihilation in the Milky Way. A\&A  \textbf{586}, A84 (2016).

\bibitem{Prantzos_2006} Prantzos N. On the intensity and spatial morphology of the 511 keV emission in the Milky Way. A\&A  \textbf{449}, 869 (2006).

\bibitem{Alexis_2014} Alexis, A.; Jean, P.; Martin, P.; Ferri\`{e}re, K. Monte Carlo modelling of the propagation and annihilation of nucleosynthesis positrons in the Galaxy.
 A\&A \textbf{564}, A108 (2014).


\bibitem{siegert_2016} Siegert, Thomas; Diehl, Roland; Greiner, Jochen; Krause, Martin G. H.; Beloborodov, Andrei M.; Bel, Marion Cadolle; Guglielmetti, Fabrizia; Rodriguez, Jerome; Strong, Andrew W.; Zhang, X. Positron annihilation signatures associated with the outburst of the microquasar V404 Cygni. Nature \textbf{531}, 341 (2016).

\bibitem{parizot_2005} Parizot, E.; Cass\^{e}, M.; Lehoucq, R.; Paul, J. GRBs and the 511 keV emission of the Galactic bulge. A\&A  \textbf{432}, 889 (2005).


\bibitem{cheng_2006} Cheng, K. S.; Chernyshov, D. O.; Dogiel, V. A. Annihilation Emission from the Galactic Black Hole. ApJ \textbf{645}, 1138 (2006).


\bibitem{totani_2006} Totani T. A RIAF Interpretation for the Past Higher Activity of the Galactic Center Black Hole and the 511 keV Annihilation Emission. PASJ \textbf{58}, 965 (2006).

	
\bibitem{Pshirkov_2016} Pshirkov M.
	Positron excess in the center of the Milky Way from short-lived $\beta^+$ emitting isotopes. Eprint arXiv:1608.06324 (2016).


\bibitem{Ascasibar_2006} Ascasibar, Y.; Jean, P.; B{\oe}hm, C.; Kn\"{o}dlseder, J. Constraints on dark matter and the shape of the Milky Way dark halo from the 511-keV line. MNRAS \textbf{368}, 1695 (2006).


\bibitem{Share_2003} Share G.H., Murphy R.J., Skibo J.G., Smith D.M., Hudson H.S., Lin R.P., Shih A.Y.,
   Dennis B.R. , Schwartz R.A., Kozlovsky B.
   High-resolution observation of the solar positron-electron annihilation line.
   ApJ \textbf{595}, L85 (2003).


\bibitem{Fleishman_2013} Fleishman, Gregory D.; Altyntsev, Alexander T.; Meshalkina, Nataliia S. Microwave Signature of Relativistic Positrons in Solar Flares. Publications of the Astronomical Society of Japan. \textbf{65}, No.SP1, article id.S7  (2013).

\bibitem{blp71} Berestetskii V.B., Lifshitz E.M., Pitaevskii L.P. Relativistic Quantum Theory. Vol. 4,  Pergamon Press (1971).


\bibitem{scord} Esposito P.B., Edenhofer P., Lueneburg E.
    Solar Corona Electron Density Distribution.
    Journal  of  Geophysical  Research \textbf{85}, 3414 (1980).

\bibitem{Schrijver} Schrijver, C. J.; Beer, J.; Baltensperger, U.; Cliver, E. W.; G\"{u}del, M.; Hudson, H. S.; McCracken, K. G.; Osten, R. A.; Peter, T.; Soderblom, D. R.; Usoskin, I. G.; Wolff, E. W. Estimating the frequency of extremely energetic solar events, based on solar, stellar, lunar, and terrestrial records. Journal of Geophysical Research: Space Physics,  \textbf{117}, Issue A8, CiteID A08103 (2012).


\bibitem{Kitze_2014} 	Kitze, M.; Neuh\"{a}user, R.; Hambaryan, V.; Ginski, C. Superflares on the slowly rotating solar-type stars KIC10524994 and KIC07133671? MNRAS,  \textbf{442}, Issue 4, 3769 (2014).

\bibitem{Candelaresi_2014} Candelaresi, S.; Hillier, A.; Maehara, H.; Brandenburg, A.; Shibata, K. Superflare Occurrence and Energies on G-, K-, and M-type Dwarfs.	
	ApJ,  \textbf{792}, Issue 1, article id. 67,  (2014)

\bibitem{Wu_2015} Wu, Chi-Ju; Ip, Wing-Huen; Huang, Li-Ching. A Study of Variability in the Frequency Distributions of the Superflares of G-type Stars Observed by the Kepler Mission. ApJ,  \textbf{798}, Issue 2, article id. 92 (2015).


\bibitem{DG_CVn} Drake, S.; Osten, R.; Page, K. L.; Kennea, J. A.; Oates, S. R.; Krimm, H.; Gehrels, N. Swift Detection of a Superflare from DG CVn. The Astronomer's Telegram, No.6121 (2014).


\bibitem{Valenti_2015}  Valenti, E.; Zoccali, M.; Gonzalez, O. A.; Minniti, D.; Alonso-García, J.; Marchetti, E.; Hempel, M.; Renzini, A.; Rejkuba, M. Stellar density profile and mass of the Milky Way bulge from VVV data. A\&A \textbf{587}, id.L6 (2015).


\bibitem{Ven_2006} van de Ven, G.; van den Bosch, R. C. E.; Verolme, E. K.; de Zeeuw, P. T. The dynamical distance and intrinsic structure of the globular cluster $\omega$ Centauri.  A\&A \textbf{445}, Issue 2,  513 (2006).







\end{thebibliography}
\end{document}